\documentclass[prl,aps,twocolumn,amssymb,showpacs]{revtex4}
\usepackage{amsmath,epsfig}
\usepackage{graphicx}
\usepackage{textcomp}

\begin{document}
\title{Determining the stability of genetic switches: explicitly accounting for mRNA noise}
\author{Michael Assaf$^{1}$, Elijah Roberts$^2$ and Zaida Luthey-Schulten$^{1,2}$}
\affiliation{Departments of $\;^1\!$Physics and $\;^2\!$Chemistry, University of Illinois at Urbana-Champaign, Urbana, IL 61801, USA} 
\begin{abstract}
Cells use genetic switches to shift between alternate gene expression states, {\it e.g.}, to adapt to new environments or to follow a developmental pathway. Here, we study the dynamics of switching in a generic-feedback on/off switch. Unlike protein-only models, we explicitly account for stochastic fluctuations of mRNA, which have a dramatic impact on switch dynamics. Employing the WKB theory to treat the underlying chemical master equations, we obtain accurate results for the quasi-stationary distributions of mRNA and protein copy numbers and for the mean switching time, starting from either state. Our analytical results agree well with Monte Carlo simulations. Importantly, one can use the approach to study the effect of varying biological parameters on switch stability.
\end{abstract}
\maketitle

Genetic switches allow cells to switch between distinct gene expression states in response to environmental stimuli and/or internal signals. The ultimate stability of these states is determined by stochastic fluctuations of mRNA and proteins during gene expression~\cite{SGE} that can give rise to spontaneous switching, even in the absence of a driving signal. When gene expression states are stable on the time scale of cellular division they can carry epigenetic information across generations, however, when they are more transient they may provide a beneficial source of heterogeneity in genetically identical populations.

Previous studies of noise-driven genetic switches have shown that switching can be treated as a first-passage problem of the underlying Markov process. These results were obtained either by using the probability generating function formalism, or by employing semi-classical approximation schemes to the chemical master equations (CMEs) or related Langevin equations~\cite{switchingbackground,Roma,wolynes,wolde}. These studies, however, focused on protein-only models and ignored the presence of mRNA and thereby the influence of transcriptional noise. Recently it has been shown that explicitly accounting for mRNA in the underlying CMEs has a strong impact on switching times~\cite{wingreen,golding}. A general method for accurately determining the mRNA/protein distributions and stability of feedback-based switches is of great interest, as they regulate diverse biological phenomena, such as microbial environmental adaptation, developmental pathways, and bacteriophage lysogeny~\cite{golding,oudenaarden,roberts}.

In this study, we explicitly account for the mRNA noise and present a concise analytical framework for accurately calculating the stability of gene expression switches subject to stochastic fluctuations of protein/mRNA. Our approach is demonstrated on a two-state positive feedback switch, which was experimentally shown to describe biological switching~\cite{xie}.
We apply a WKB theory~\cite{Bender} to the CMEs and obtain the quasistationary probability distributions of mRNA and protein copy numbers in the on and off states, from which we extract the mean switching times starting from either state. Our results agree well with Monte Carlo simulations. Finally, we use our analytical predictions to study the effect of promoter fluctuations on switching stability.

\begin{figure}
\includegraphics{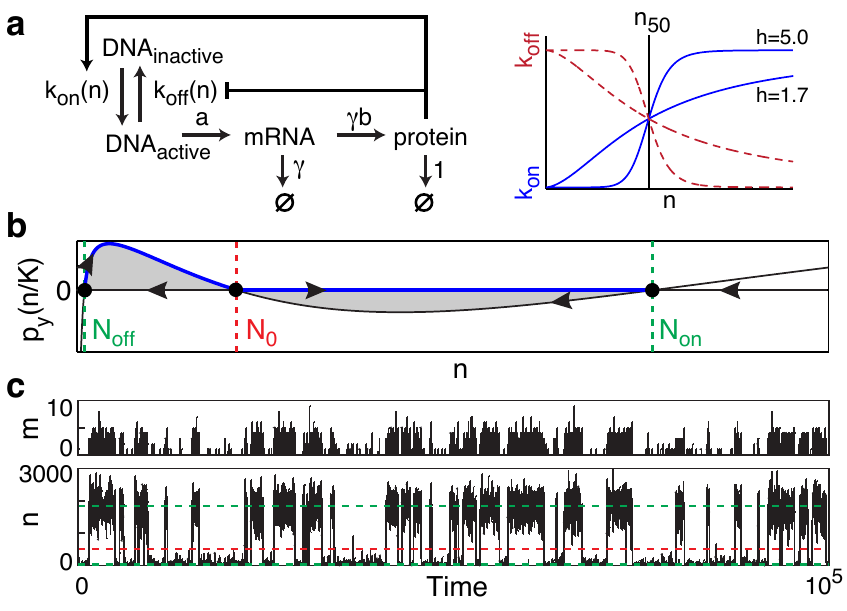}
\caption{(Color online) (a) Model for positive feedback network. Transcription and translation are modeled as first-order processes with rates $a$ and $\gamma b$, respectively. mRNA and proteins undergo first order degradation with rates $\gamma$ and $1$ (we rescale rates by the protein decay rate). The feedback functions $k_{on}(n)$ and $k_{off}(n)$ control promoter transitions. (b) The momentum $p_y$ vs protein copy number $n$. The thick line indicates the $off{\to}on$ switching trajectory, and shaded areas correspond to the entropy barriers for switching. (c) mRNA $m$ and protein counts in a typical Monte Carlo trajectory undergoing switching. In (b,c) $K\!=\!ab\!=\!2400$, $b\!=\!22.5$, $h\!=\!2$, $n_{50}\!=\!1000$, $k_0^{min}\!=\!k_1^{min}\!=\!a/100$, $k_0^{max}\!=\!k_1^{min}\!=\!a$, and $\gamma\!=\!50$.}
\label{fig:fig1}
\end{figure}

We consider a two-state gene-expression model where transitions between a transcriptionally active and inactive promoter are controlled by the protein copy number $n$ via positive feedback (see Fig.~\ref{fig:fig1}(a)). The transition rates into the active and inactive states are $k_{on}(n)\!\equiv\! f(n)$ and $k_{off}(n)\!\equiv\! g(n)$.\,While our analytical treatment holds for generic $f(n)$ and $g(n)$, we consider a concrete example using Hill-type functions $f(n)=k_0^{min}\!+\!(k_0^{max}\!-\!k_0^{min})n^{h_1}/(n_{50}^{h_1}+n^{h_1})$ and $g(n)\!=\!k_1^{max}\!-\!(k_1^{max}\!-\!k_1^{min})n^{h_2}/(n_{50}^{h_2}+n^{h_2})$, which were shown to be biologically relevant, \textit{e.g.}, in the {\it lac} operon~\cite{roberts}. Here $n_{50}$ is the curve's midpoint and for simplicity we set $h_1=h_2=h$.

\begin{figure}
\includegraphics{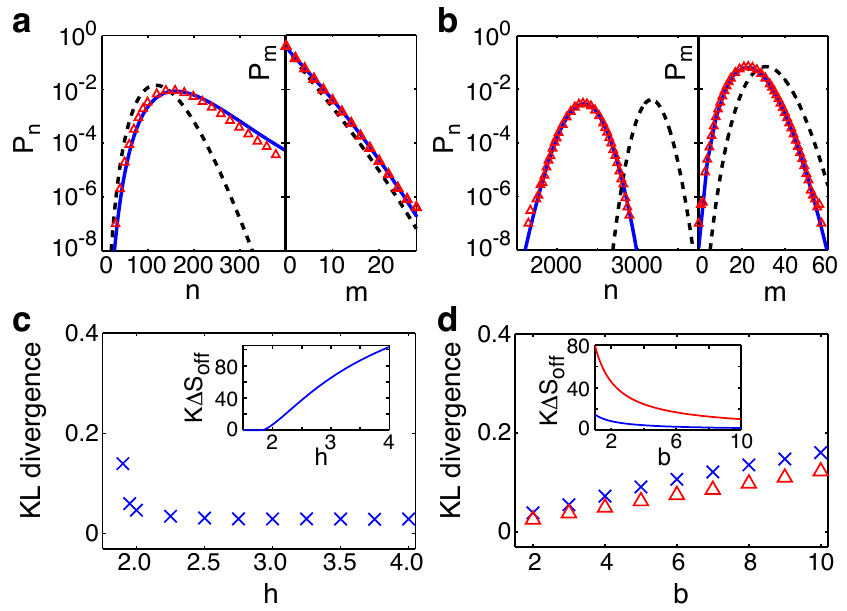}
\caption{(Color online) (a) Protein (left) and mRNA (right) QSDs in the off state showing WKB result~(\ref{pdfinactive}) (solid) and MC simulations ($\triangle$) for $b\!=\!2$ and $h\!=\!2$. Our results converge to those of~\cite{swain} (dashed) for $h\to\infty$. (b) As in (a) for the on state. (c) The Kullback-Leibler divergence (see text) vs. $h$ comparing the WKB and MC PDFs of the off state for $b\!=\!2.5$. (d) The KL divergence vs. $b$ of the off state for $h\!=\!2$ ($\times$) and $h\!=\!2.5$ ($\triangle$). Insets: The theory holds for $K\Delta S\gg 1$. Other parameters are $K\!=\!ab\!=\!3200$, $n_{50}\!=\!2000$, $k_0^{min}\!=\!a/50$, $k_0^{max}\!=\!a$,  $k_1^{min}\!=\!a/100$, $k_1^{max}\!=\!a/2$, and $\gamma\!=\!50$.}
\label{fig:fig2}
\end{figure}

The deterministic rate equations (DREs) for the \textit{mean number} of mRNA, $M$, and proteins, $N$, read
\begin{equation}
\dot{M}=a\, f(N)/[f(N)+g(N)] - \gamma M,\;\;\;\dot{N}=\gamma b M-N,\label{meanfield}
\end{equation}
where $f(N)/[f(N)+g(N)]$ is the probability for an active promoter. To exhibit bistability, Eqs.~(\ref{meanfield}) must have (at least) three (positive) fixed points. We denote by $N_{on}$ and $N_{off}$, respectively, the attracting fixed points corresponding to the average protein copy number in the on and off states, and assume that $1\ll N_{off}\ll N_{on}$. These points are separated by a repeller $N_0$ such that  $N_{off}<N_0<N_{on}$. For biologically-relevant parameters, see below, one has $a\sim k_{0,1}^{max}\gg k_{0,1}^{min}$, and $b={\cal O}(1)$. Also, when $h\gg 1$, $N_{on}\simeq a\,b\simeq N_{off}k_1^{max}/k_0^{min}$. Thus, $K\equiv a\,b\gg 1$ -- the typical protein number in the on state -- will serve as the large parameter of the theory.

DREs~(\ref{meanfield}) ignore noise and predict that, once the system has settled in one of the attracting fixed points, it stays there forever. Yet, the presence of intrinsic noise allows switching between these fixed points by crossing the corresponding entropy barrier~\cite{Dykman}.  In the stochastic picture, starting from the vicinity of either state the system rapidly converges into the quasistationary distribution (QSD) about this state. This distribution is metastable, and slowly decays due to a (exponentially small) \textit{probability leakage} through the entropy barrier at $N_0$~\cite{Dykman,AM}. It is this leakage that determines the corresponding switching rates between the metastable states.

To model the stochastic behavior of the switch, we use two coupled CMEs. These describe the dynamics of $P_{m,n}$ and $Q_{m,n}$ -- the  probability distribution functions (PDFs) of having $m$ mRNAs and $n$ proteins at time $t$ with the promoter in the inactive and active state, respectively:
\begin{eqnarray}
&&\hspace{-9mm}\dot{P}_{m,n}\!=\!g(n)Q_{m,n}\!-\!f(n)P_{m,n}\!+\!\mathbf{A}P_{m,n}\nonumber\\
&&\hspace{-9mm}\dot{Q}_{m,n}\!=\!-g(n)Q_{m,n}\!+\!f(n)P_{m,n}\!+\!\left[\!\mathbf{A}\!+\!a(E_m^{-\!1}\!-\!1)\right]\!Q_{m,n}.
\label{fullmaster}
\end{eqnarray}
Here, $E_n^{j}f(n)=f(n+j)$, $\mathbf{A}\equiv (E_n^{1}-1)n+\gamma(E_m^1-1)m+\gamma b m(E_n^{-1}-1)$ is a birth-death operator related to the inactive promoter, and $\sum P_{m,n}+Q_{m,n}=1$. We are seeking the QSD starting from the vicinity of $N_{off}$ (the on state treatment is equivalent, see below). Putting $\dot{P}_{m,n}=\dot{Q}_{m,n}=0$ in Eqs.~(\ref{fullmaster}) (that are exponentially small for $K\gg 1$), and eliminating, \textit{e.g.}, $Q_{m,n}$ we obtain
\begin{equation}
0\!=\!\left\{\mathbf{A}+g(n)^{-1}\left[\mathbf{A}+a(E_m^{-1}\!-\!1)\right]\![f(n)\!-\!\mathbf{A}]\right\}P_{m,n}.
\label{master1}
\end{equation}

In the case of \textit{constant} transition rates between the active and inactive states, Eqs.~(\ref{master1}) were asymptotically solved in the $\gamma\gg 1$ limit using the probability generating function~\cite{swain,Gardiner}. However, for generic feedback functions the generating function
formalism cannot be used.

Instead, we use here a powerful method based on the WKB approximation~\cite{Bender}, to treat the (quasi)stationary CMEs~(\ref{master1})~\cite{Dykman}. The WKB ansatz reads
\begin{equation}\label{wkb}
P_{m,n}\equiv P(x,y)\sim \exp[-KS(x,y)].
\end{equation}
Here $x=m/K$ and $y=n/K$ are the densities of the mRNA and proteins, respectively. Plugging ansatz~(\ref{wkb}) into Eq.~(\ref{master1}), \textit{e.g.} the step operator $E_m^{\pm 1}$ is replaced in the leading order by the function $e^{\mp \partial_x S(x,y)}$. After some algebra, one arrives in the leading order at a stationary Hamilton-Jacobi equation $H(x,y,\partial_x S,\partial_y S)=0$, with
\begin{equation}
H=\mathcal{A}+\tilde{g}(y)^{-1}\left[\mathcal{A}+b^{-1}(e^{p_x}-1)\right][\tilde{f}(y)-\mathcal{A}].
\label{Ham}
\end{equation}
Here $\mathcal{A}\!=\!\mathcal{A}(x,y,p_x,p_y)\!=\!y(e^{-p_y}\!-\!1)+\gamma x(e^{-p_x}\!-\!1)+\gamma b x(e^{p_y}\!-\!1)$ is now a function, and we have used the rescaled feedback functions $\tilde{f}(y)=f(y)/K$ and $\tilde{g}(y)=g(y)/K$, and mRNA production rate $a/K=b^{-1}$. Also, in analogy to classical mechanics we have introduced the momenta $p_x=\partial_x S(x,y)$ and $p_y=\partial_y S(x,y)$, corresponding to the steepness of the sought PDF's~\cite{Dykman,EKAM}. Note that putting $p_x=p_y=0$ corresponds to mean-field dynamics; in this case using~(\ref{Ham}) the Hamilton's equations for $\dot{x}=\partial_{p_x}H$ and $\dot{y}=\partial_{p_y}H$ become DREs~(\ref{meanfield}) divided by $g(n)/[f(n)+g(n)]$ [since~({\ref{Ham}) corresponds to the inactive state occupied with probability $g(n)/[f(n)+g(n)]$].

The strength of this theory is that it can accurately account for \textit{rare large fluctuations} responsible for switching. To do so one has
to solve the Hamilton's equations for $\dot{x}$ and $\dot{y}$ together with $\dot{p_x}=-\partial_{x}H$ and $\dot{p_y}=-\partial_{y}H$. Now, as we look for a zero-energy trajectory of~(\ref{Ham}), the action reads $S(x,y)=\int p_x dx+p_y dy$, which yields PDF~(\ref{wkb}). 2\-D Hamiltonian systems can be solved numerically~\cite{Roma}. In our case such a solution would yield the complete statistics for arbitrary mRNA degradation rates, which is important {\it e.g.} in eukaryotic systems where $\gamma = {\cal O}(1)$.

Further analytical progress can be made in the regime of $\gamma\gg 1$ (relevant for bacterial systems), for which the mRNA dynamics is enslaved to that of the protein~\cite{swain}. We adiabatically eliminate~\cite{PAM} the fast component in the mRNA dynamics by assuming $x$ and $p_x$ rapidly converge to slowly-varying functions of $(y,p_y)$. Taking $y$ and $p_y$ constant and putting $\dot{x}=\dot{p_x}=0$, one obtains $x={\cal O}(\gamma^{-1})$~\cite{xdep}, and $p_x=-ln(1+b-be^{p_y})$~\cite{px}, so that Eq.~(\ref{Ham}) becomes a \textit{reduced} Hamiltonian $H_r(y,p_y)$
\begin{eqnarray}\label{hred}
H_r\!=\!(z^{-1}\!-\!1)\!\left\{\!y\!+\!\left[\!y\!+\!\frac{z}{b(z\!-\!1)\!-\!1}\right]\!
\!\!\left[\frac{\tilde{f}(y)\!-\!y(z^{-1}\!-\!1)}{\tilde{g}(y)}\!\right]\!\!\right\}
\end{eqnarray}
with $z\equiv e^{p_y}$. This Hamiltonian effectively accounts for the fact that the proteins are produced in geometrically distributed bursts with mean $b$, which in turn asymptotically accounts for the mRNA noise when $\gamma\gg 1$. However, had one initially eliminated the mRNA species from CMEs~(\ref{fullmaster}) using geometrically distributed protein births, one would have obtained an analytically intractable one-dimensional CME. This is because the system under consideration is a two-state switch with non-linear feedback.

The (nontrivial) \textit{zero-energy} trajectory of Hamiltonian~(\ref{hred}) encodes the stochastic dynamics  of (only) the proteins, and corresponds to its quasi-stationary behavior. This trajectory gives $p_y$  as function of $y$ and represents the most probable path the stochastic system follows while undergoing switching~\cite{Dykman,EKAM}. The normalizable solution reads $p_y(y)=\ln [(-B+\sqrt{B^2-4AC})/(2A)]$, see Fig.~\ref{fig:fig1}(b), where $A=(1+by)[y+\tilde{f}(y)]+by\tilde{g}(y)$, $B=-y[y(1+2b)+1+(1+b)(\tilde{f}(y)+\tilde{g}(y))]$, and $C=(1+b)y^2$. Thus $S(y)\simeq \int^y p_y(y')dy'$~\cite{pxdx}, and using~(\ref{wkb}) we have $P(y)\sim e^{-KS(y)}$. Note that $P(y)$ is the contribution to the QSD corresponding to an \textit{inactive} promoter. A similar contribution from the \textit{active} promoter can be shown to also satisfy $Q(y)\sim e^{-KS(y)}$ using~(\ref{fullmaster}). Thus, the protein copy number QSD, ${\cal P}_n$, starting from the vicinity of the off state, reads ${\cal P}_n\equiv {\cal P}(y)=P(y)+Q(y)\sim e^{-KS(y)}$. Expanding $S(y)$ in the vicinity of $y_{off}=N_{off}/K$ up to second order, and demanding that the Gaussian integration be normalized to $1$, the \textit{normalized} ${\cal P}(y)$ satisfies
\begin{equation}\label{pdfinactive}
{\cal P}(y) \simeq \sqrt{S''(y_{off})/(2\pi K)}\,e^{-K[S(y)-S(y_{off})]}.
\end{equation}
Note that the preexponent entering~(\ref{pdfinactive}) holds only in the Gaussian regime of the PDF, whose width is $\sigma=\sqrt{K/S''(y_{off})}$. One can check that the on state QSD coincides with~(\ref{pdfinactive}) upon replacing $y_{off}\!\to\!  y_{on}$. The Gaussian normalization above is valid when the QSD's width is sufficiently small compared to $N_{off}$ for the off state (and $N_{on}-N_0$ for the on state). From~(\ref{pdfinactive}) one can readily find the joint QSD, ${\cal P}_{m,n}={\cal P}_{m|n}{\cal P}_n$, where ${\cal P}_{m|n}$, the probability to find $m$ mRNA molecules \textit{given} $n$ proteins, can be found using standard techniques~\cite{swain}. Given ${\cal P}_{m,n}$, the mRNA QSD satisfies ${\cal P}_m=\sum_n {\cal P}_{m,n}$.

Our result~(\ref{pdfinactive}) can be compared to that of Ref.~\cite{swain} \textit{e.g.} for the one-state model, where the promoter is always active. Putting $\tilde{f}(y)=1$ and $\tilde{g}(y)=0$, the momentum becomes $p_y(y)\!=\!\ln[((1\!+\!b)y)/(1\!+\!by))]$ so that Eq.~(\ref{pdfinactive}) simplifies to
${\cal P}_n\!=\![2\pi ab(b\!+\!1)]^{-1/2}a^{-a}n^{-n}(a\!+\!n)^{a+n}b^n (1\!+\!b)^{-(n+a)}.$ This result coincides with the $n\gg 1$ asymptote of Eq.~(9) in~\cite{swain} by using the Stirling formula. As expected, the prefactor here coincides with that of~\cite{swain} only in the Gaussian region of the fixed point $n\!=\!ab$.

To check our theoretical predictions for \textit{generic non-constant} $f(n)$ and $g(n)$, we performed Monte Carlo (MC) simulations using the Gillespie algorithm~\cite{gillespie}. An example of a typical MC run can be seen in Fig.~\ref{fig:fig1}(c). In Fig.~\ref{fig:fig2}(a,b) we compare the WKB prediction (where the action is found by numerical integration) for the protein and mRNA QSDs for the off (a) and the on (b) states, with MC simulations and results of Ref.~\cite{swain}. The latter are expected to be valid only in the limit of $h\gg 1$ (when the feedback functions become approximately step functions). In panels (c,d)  we show the Kullback-Leibler (KL) divergence $\sum P^{(1)}_n\ln(P^{(1)}_n/P^{(2)}_n)$ (a measure of the difference between PDFs $P^{(1)}_n$ and $P^{(2)}_n$) between WKB result~(\ref{pdfinactive}) and MC simulations for various parameters.

\begin{figure}
\includegraphics{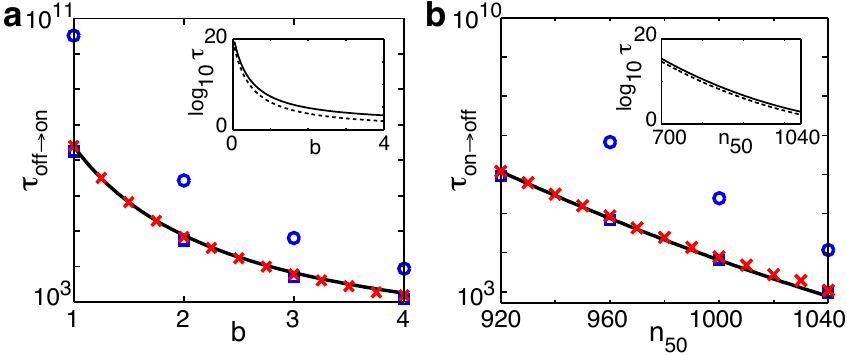}
\caption{(Color online) (a) MST $\tau_{off\to on}$ as a function of $b$ for $n_{50}\!=\!720$. WKB result~(\ref{tau}) with numerical preexponent (solid) and MC simulations ($\times$). (b) $\tau_{on\to off}$ as a function of $n_{50}$ for $b\!=\!15$. Other parameters are $h\!=\!2$, $K\!=\!ab\!=\!2400$, $k_0^{min}\!=\!k_1^{min}\!=\!a/100$, $k_0^{max}\!=\!k_1^{min}\!=\!a$ and $\gamma\!=\!50$. Preexponents were $17.8$ (a) and $4.8$ (b). Also shown are the MSTs from MC simulations of protein burst models with constant ($\bigcirc$) and geometrically distributed ($\square$) burst sizes. Insets: WKB result without (dashed) and with (solid) numerical preexponent. }
\label{fig:fig3}
\end{figure}

Now, the mean switching time (MST) $\tau_{off\to on}$ can be readily inferred from QSD~(\ref{pdfinactive}): it is the inverse of the flux through the repelling fixed point $y_0\!=\!N_0/K$~\cite{Dykman,AM}. The logarithm of the MST is proportional to the effective entropy barrier between the attracting and repelling fixed points $\Delta S_{off}=S(y_0)-S(y_{off})$. Therefore, we have
\begin{equation}\label{tau}
\ln \tau_{off\to on}=K\left[\Delta S_{off}+{\cal O}\left(1/K,1/\gamma\right)\right],
\end{equation}
with $\tau_{on\to off}$ the same using $\Delta S_{on}=S(y_0)-S(y_{on})$. Note, that while the prefactor of the MST is unknown, based on single-species calculations it is expected to be ${\cal O}(1)$ and independent of $K$~\cite{EKAM}. Eq.~(\ref{tau}) indicates the WKB formalism is valid for $K\Delta S\gg 1$ (see Fig.~\ref{fig:fig2} insets).

In Fig.~\ref{fig:fig3} we compare the theoretical MST prediction to MC simulations. Panel (a) compares $\tau_{off\to on}$ vs $b$; a nontrivial super-exponential dependence is observed. It is also shown that the functional dependence of $\tau_{off\to on}$ is excellently captured by~(\ref{tau}) with a numerical slowly-varying preexponent. Panel (b) compares $\tau_{on\to off}$ vs $n_{50}$, and again the functional dependence is excellently captured by the theoretical MST. Fig.~\ref{fig:fig3} also shows that the mRNA noise can be accounted for when $\gamma \gg 1$ by assuming geometrically distributed protein births, while using a constant burst size yields markedly different results.

Promoter fluctuations are known to have a substantial impact on the PDFs and switching times of genetic switches \cite{wolynes,wolde}. We used  Eq.~(\ref{tau}) to study switch stability with respect to the promoter transition dynamics. Multiplying $k_{on}$ and $k_{off}$ by $\alpha$ we control the frequency and duration of mRNA bursts in the off state and pauses in the on state, while leaving unchanged the relative probability of the promoter to be in the active/inactive states.

\begin{figure}
\includegraphics{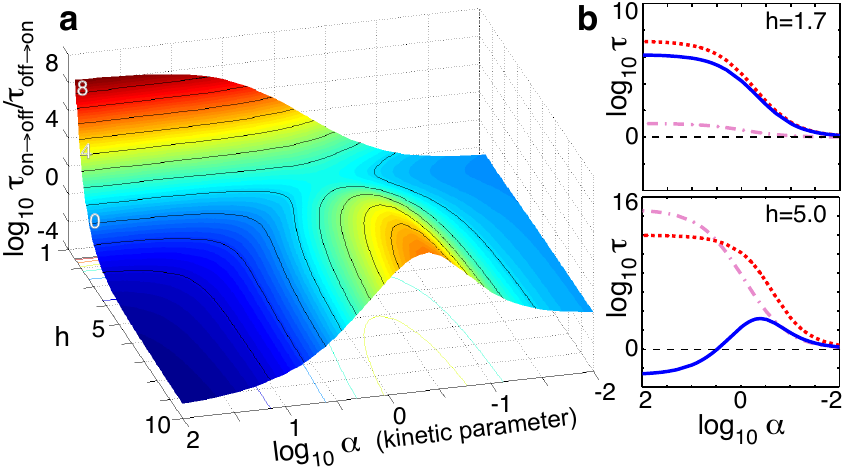}
\caption{(Color online) (a) $\tau_{on\to off}/\tau_{off\to on}$ (on state relative stability) vs $h$ and $\alpha$, for $k_0^{min}\!=\!k_1^{min}\!=\! a\alpha/100$, $k_0^{max}\!=\!k_1^{min}\!=\!a\alpha$, $K\!=\!2400$, $b\!=\!15$, and $n_{50}\!=\!950$. (b) MSTs $\tau_{on\to off}$ (dashed), $\tau_{off\to on}$ (dash-dotted), and their ratio (solid).}
\label{fig:fig4}
\end{figure}

By increasing $\alpha$ the duration of bursts and pauses diminishes, large fluctuations become rarified, and crossing the entropic barrier becomes harder (see also~\cite{wolde}). The rate of increase of the entropic barrier, however, differs for the ${off\to on}$ and ${on\to off}$ transitions, as seen in Fig.~\ref{fig:fig4}. For small $h$ the rate of increase of the ${on\to off}$ entropic barrier exceeds that of $off \to on$ for the entire range of $\alpha$, thereby amplifying the on state stability. However, this is not the case for higher $h$, which gives rise to a non-monotonic stability curve for the on state (see panels b in Fig.~\ref{fig:fig4}). These results stress the role of promoter kinetics, not just thermodynamics, for genetic switches, which are inherently far from equilibrium.

We have presented an analytical framework for the accurate analysis of genetic switches while explicitly accounting for mRNA noise. This framework is expected to be useful for studying diverse genetic circuits characterized by metastable switching, \textit{e.g.}, those with additional promoter states such as DNA looping or nucleosome remodeling. In particular, it can be used to help elucidate the underlying regulatory circuits responsible for phenotypical changes as a result of switching.

M.\,A. acknowledges the Rothschild and Fulbright foundations for support. E.\,R. and Z.\,L.\,S acknowledge support from the DOE Office of Science (BER), and from the NSF via the CPLC at UIUC (PHY-0822613).

\newcommand{\bibfnamefont} [1] {#1}
\newcommand{\bibnamefont} [1] {#1}


\begin{thebibliography}{99}
\bibitem{xdep} $x(y,p_y)$ is not given explicitly, since plugging $p_x(p_y)$ into Hamiltonian~(\ref{Ham}) eliminates the $x$-dependence.
\bibitem{px} This relation is analogous to that of~\cite{swain} between $u$\,and\,$v$.
\bibitem{pxdx} Here we have neglected in $S(y)$ the term $\int p_x dx\sim {\cal O}(\!\gamma^{-1}\!)$\,[which can be written as function of $y$ using $p_y(y)$].
\end{thebibliography}
\end{document}